\documentclass[twocolumn,floatfix,showkeys,showpacs,preprintnumbers,amsmath,amssymb]{revtex4}

\usepackage{graphicx}
\usepackage{subfigure}
\usepackage{dcolumn}
\usepackage{bm}
\begin{document}

\title{Accelerator Based Production of $^{225}$Ac for Cancer Treatment\\}
\author{H. Kumawat$^{1,2}$\footnote{author. Email address: harphool@barc.gov.in}}
\author{S. V. Suryanarayana$^{1}$}%
\affiliation{$^1$Nuclear Physics Division, BARC, Mumbai-400085, India\\}
\affiliation{$^2$Homi Bhabha National Institute, Anushaktinagar, Mumbai 400094, India\\}
\date{\today}

\begin{abstract}
Targeted Alpha Therapy potentially offers a more specific action in killing tumor cells and less damage to neighbouring normal cells compared to $\beta$-emitters. $^{225}$Ac along with three other $\alpha$-emitting daughter nuclei $^{221}$Fr, $^{217}$At and $^{213}$Po have significant potency for clinical use. Monte carlo simulation has been carried out using MONC code for production routes of $^{225}$Ac for its possible generation using proton accelerators with Thorium-232 and Radium-226 targets. All the calculations has been carried out using global model parameters without adjusting for individual target. The study reveals that proton beam of 20-30MeV energy with 5$\mu$A current on radium-226 target can give rise to 1Ci of $^{225}$Ac activity in a week of irradiation. Thorium irradiation at low proton energy produces smaller $^{225}$Ac activity but one can benefit from high current, stable proton accelerators.  Proton energy of 100MeV or above with hundreds of $\mu$A current can produce reasonable $^{225}$Ac activity for practical use. However, the contamination of other co-produced radio-isotopes is much less at low proton energy hence low energy high current proton accelerators are also viable alternative to generate Actinium-225 radio isotope. 
\end{abstract}

\keywords{$^{225}$Ac, $^{232}$Th, $^{226}$Ra, Targeted Alpha Therapy, proton irradiation}

\maketitle
\section{\label{sec:level1}Introduction}
Radio-isotope production by different methods, chemical seperation, isotope delivery, molecular targets, radiolabeling protocols and identification of clinical applications are an active area of research. Several $\beta$-emitters, $^{90}$Y (t$_{1/2}$=2.67d) and $^{131}$I (t$_{1/2}$=8.07d) emit electrons with maximum kinetic energies of 0.3–2.3 MeV with a range of $\sim$0.5 - 12 mm in tissues. Single cell disease such as leukemia, micrometastases, post-surgical residual disease and some other types of cancer may not be curable with targeted $\beta$-particle therapy due to insufficient dose to each cell \cite{humm90}. Concomitantly, a significant portion of the dose would also be deposited in the surrounding normal tissues by virtue of long range. Therefore $\alpha$-particle targeted therapy provides superior option compared to $\beta$-particle therapy. Production of $^{225}$Ac has been studied for its possible generation using proton accelerator.

Heavy ions have larger linear energy transfer and smaller penetration depth which lead to targeted killing of the cancer cell. Carbon therapy is becoming popular after proton therapy due to larger linear energy transfer. Targeted Alpha Therapy (TAT) is one of the promising and effective methods for treatment of bladder cancer, brain tumors, neuroendocrine tumors and prostate cancer due to short range (47-85$\mu$m) \cite{sgouros1,essler12} of alpha particles. The linear energy transfer from alpha particles is two orders of magnitude higher than that by $\beta$-particle. Single $\alpha$-emitter such as $^{213}$Bi eluted from $^{225}$Ac or $^{225}$Ac with 4-$\alpha$ emitting isotopes along with radio-isotopes from its decay chain are currently under immense interest for treating several malignant diseases even acute myeloid leukemia. $^{225}$Ac and daughter isotopes with 4-$\alpha$'s give 1000 times more integrated dose compared to single alpha from $^{213}$Bi\cite{Brechbiel07}. Relative biological effectiveness is 3-8 times higher for alpha particles \cite{Azure94}. Although, $^{225}$Ac is superior $\alpha$-emitting source but the fate of the free daughter radioisotopes in circulation after decay is to be resolved thoroughly.

Alpha emitters such as $^{225}$Ac are favorable due to their very short range (few cell diameters) and targeted high doses to kill cancer cells which show resistance to treatment with beta- or gamma- irradiation or chemotherapeutic drugs. Some of the gamma emission ($^{221}$Fr, 218.2keV, 11.6$\%$; $^{213}$Bi, 440.5keV, 26.1$\%$; $^{209}$Tl, 117.2keV, 84.3$\%$, 465.1keV, 96.9$\%$, 1567.1keV, 99.8$\%$;) from the decay chain of the $^{225}$Ac along its path gives $\textit{in vivo}$ imaging and pharmaco-kinetic and dosimetric studies \cite{McDevitt01}.  Long half-life  of $^{225}$Ac and the multiple alpha particles generated in the decay chain render $^{225}$Ac a particularly cyto-toxic radio-nuclide. Reactor based production of $^{225}$Ac by decay of $^{229}$Th (T$_{1/2}$=7880y) is available from few places like ORNL, USA, IPPE, Obninsk, Russia and ITU, Germany with combined inventory of $\sim$1.8Ci/year\cite{iaea2013}. There are plan to develop advance reactor system with $^{232}$Th utilization \cite{shina06}. This reactor can generate $^{229}$Th which can give $^{225}$Ac through $\alpha$-decay.

Several studies have been conducted to investigate different means of increasing the available supply of either $^{229}$Th parent or $^{225}$Ac itself.  $^{225}$Ac can be produced by proton+$^{232}$Th reaction at high energy spallation reaction as a direct product as well as it can be milked from $^{229}$Pa at low energies. $^{229}$Pa decays via emitting $\alpha$-particle to $^{225}$Ac with only 0.48$\%$ branching ratio and rest decays to $^{229}$Th. Thus, producing required quantities of $^{225}$Ac activity using low energy proton accelerator from $^{229}$Th is challenging due to its extremely long half-life despite the $\sim$160 mb cross section for the $^{232}$Th(p, 4n)$^{229}$Pa reaction at $\sim$30 MeV. Dedicated accelerator production of $^{229}$Th may not be economically viable due to long irradiation times and high currents required to produce a substantial quantity of $^{229}$Th \cite{jost2013}. 

Cross-section were measured at ITU, Germany for $^{226}$Ra(p, 2n)$^{225}$Ac to seek direct production of this isotope \cite{Apostolidis05}. The cross-section peaks ($\sim$ 700mb) around 17MeV proton energy. A feasibility study of the $^{226}$Ra($\gamma$, 2n)$^{225}$Ra reaction for producing the $^{225}$Ra, the parent to $^{225}$Ac, has revealed that $^{225}$Ra yields are insignificant for practical use \cite{Melville07}. 

In this paper, activity of $^{225}$Ac is estimated for p+$^{232}$Th and p+$^{226}$Ra reactions at low proton energies. A comparative study for the production of $^{225}$Ac by Thorium-232 target at different energies is performed to search an optimum energy. Monte-carlo code MONC ($\textbf{\underline{MO}}$nte carlo $\textbf{\underline{N}}$ucleon transport $\textbf{\underline{C}}$ode) \cite{hkumawat05,hkumawat04,hkroot,hkvenkata13,baras1} is used for the present simulation. In Sec. II, we present small description of MONC. Section III contains simulation studies for production of $^{225}$Ac. Conclusions are given in Sec. IV. 

\section{\label{sec:level2} Description of MONC}
Monte Carlo program MONC incorporates Intra-nuclear Cascade, Pre-equilibrium, 
Evaporation and Fission models to simulate spallation reaction mechanism for 
thin and thick targets. Modeling details of Intra-nuclear cascade, pre-equilibrium 
particle emission are described in detail in Ref. \cite{baras1}. Treatment of cutoff energy 
from Intra-nuclear to pre-equilibrium and next to evaporation stage was changed 
later which is described in Ref. \cite{hkumawat05}. Generalized evaporation model was developed 
as described in Ref. \cite{mashnik}. Fission barrier, level density parameter and 
inverse cross sections for pre-equilibrium/evaporation/fission model are given in detail in Ref. \cite{hkumawat05, hkumawat04}. The following level density parameters are used in MONC.
\begin{equation}
 a =  (0.134-.000121A)(1+(1-e^{-0.061E})\Delta S/E)     \label{eqpa229}
\end{equation}
Where $\Delta$S is shell correction, E is excitation energy and A is Mass number of the compound nucleus.
Benchmark of spallation models for experimental values of neutron, charged particles, 
and pions double differential production cross-sections, particle multiplicities, 
spallation residues and excitation functions was organized by IAEA and is given in Ref. \cite{hkumawat10}. 
We have used the predecessor of this code named CASCADE.04 to calculate these quantities in 
the benchmark. Heat Deposition algorithm for thick spallation targets and thin films was
modified and benchmarked as mentioned in Ref. \cite{hkheat08}. The code was further developed for 
the Neutron shielding and dosimetry applications \cite{hkumawat09}. The code can be used for single nucleus interaction in basic reaction studies and can be invoked for the thick target simulation during the transport. Energy loss of the charged particle is calculated during the transport in the thick target.

Point data is used for the low energy neutron transport as described in \cite{hkvenkata13,hkroot}. The processing of ENDF/B neutron data library and generation of point data at required temperature is inherent feature of the code. Linear chain method is implemented to solve the Bateman's\cite{bateman06} general solution for decay and build-up of isotope inventories. ENDF-decay data library is used for the decay properties of different radio-active isotopes.

\section{\label{sec:level3}Production of Actinium}

\begin{figure}
\includegraphics[scale=0.5]{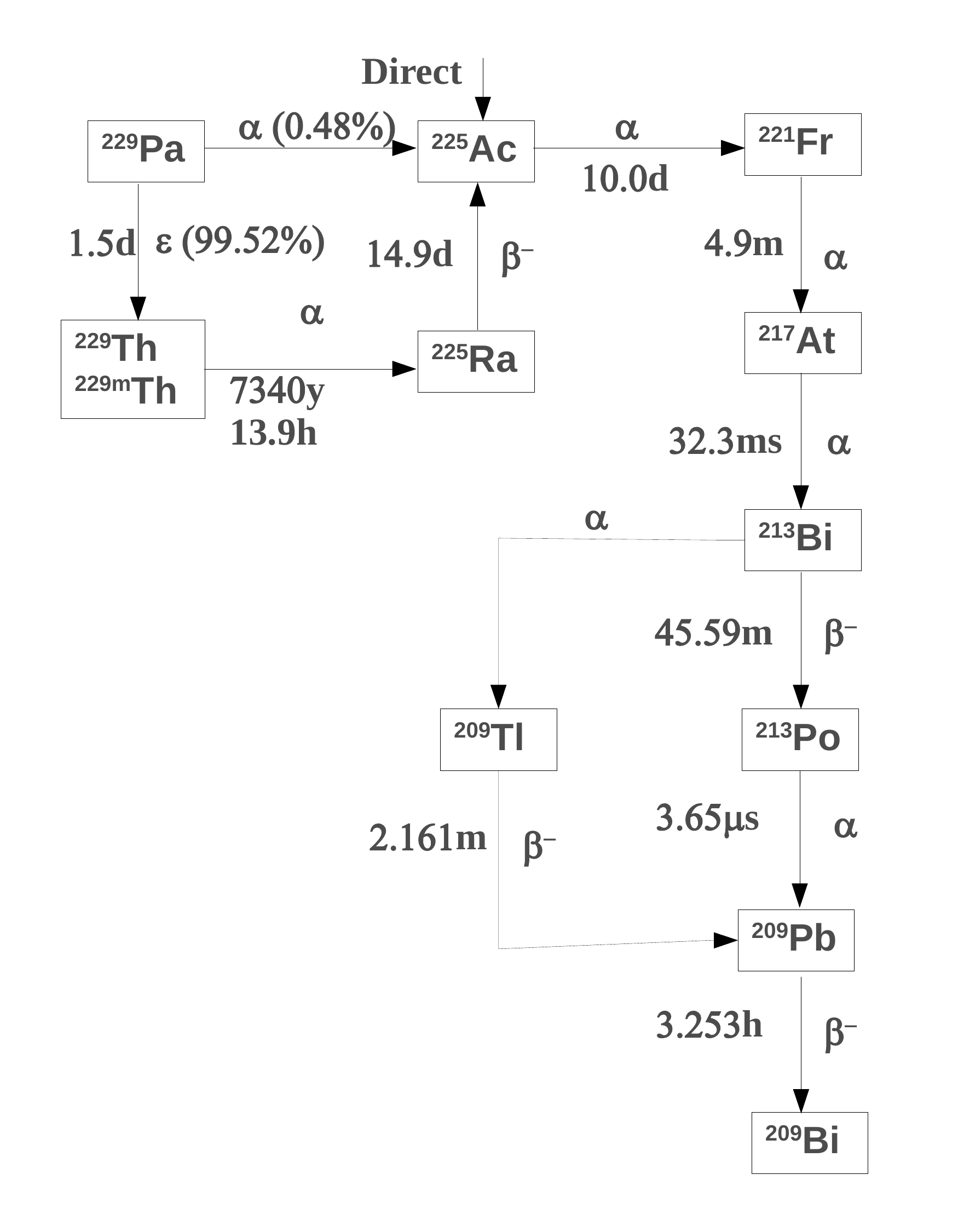}
\caption{Decay chain of $^{225}$Ac.} \label{acchain}
\end{figure}

The production of $^{225}$Ac at low energy is governed by generation of $^{229}$Pa through $^{232}$Th(p, 4n)$^{229}$Pa reaction. The reaction for  production of $^{229}$Pa as given in Eq. \ref{eqpa229}, has a threshold of $\sim$19.5MeV. Direct production of $^{225}$Ac starts at higher energies. The cross-section for direct production of $^{225}$Ac at low energies is rather low but it is generated by decay of $^{229}$Pa through direct $\alpha$ decay with 0.48$\%$ branching ratio; rest of it decays through $\beta^-$ decay to $^{229}$Th. Full decay chain is given in Fig. \ref{acchain}. The decay products of the $^{225}$Ac are also $\alpha$-emitters untill it gives $^{209}$Bi as stable isotope.
\begin{equation}
 p + ^{232}Th \Rightarrow 4n + ^{229}Pa  (\Delta Q = -19.446 MeV) \label{eqpa229}
\end{equation}
It is clear from Fig. \ref{acchain} that $\alpha$ decay branching ratio leading to production of $^{225}$Ac in a short span of time is only 0.48$\%$ and remaining 99.52$\%$ goes to $^{229}$Th which has long half life of 7340 years and is also available from weapons program or reactors. In case of reactor based production from Thorium, the route is given as

\begin{equation}
\begin{aligned}
{} & \{n + ^{232}Th \Rightarrow ^{233}Th \Rightarrow ^{233}U +\beta^- \\
      & \Rightarrow^{229}Th+\alpha\Rightarrow^{225}Ra+\alpha \Rightarrow^{225}Ac+\beta^-\}\label{eqra229}
\end{aligned}
\end{equation}

\begin{figure}
\includegraphics[scale=0.6]{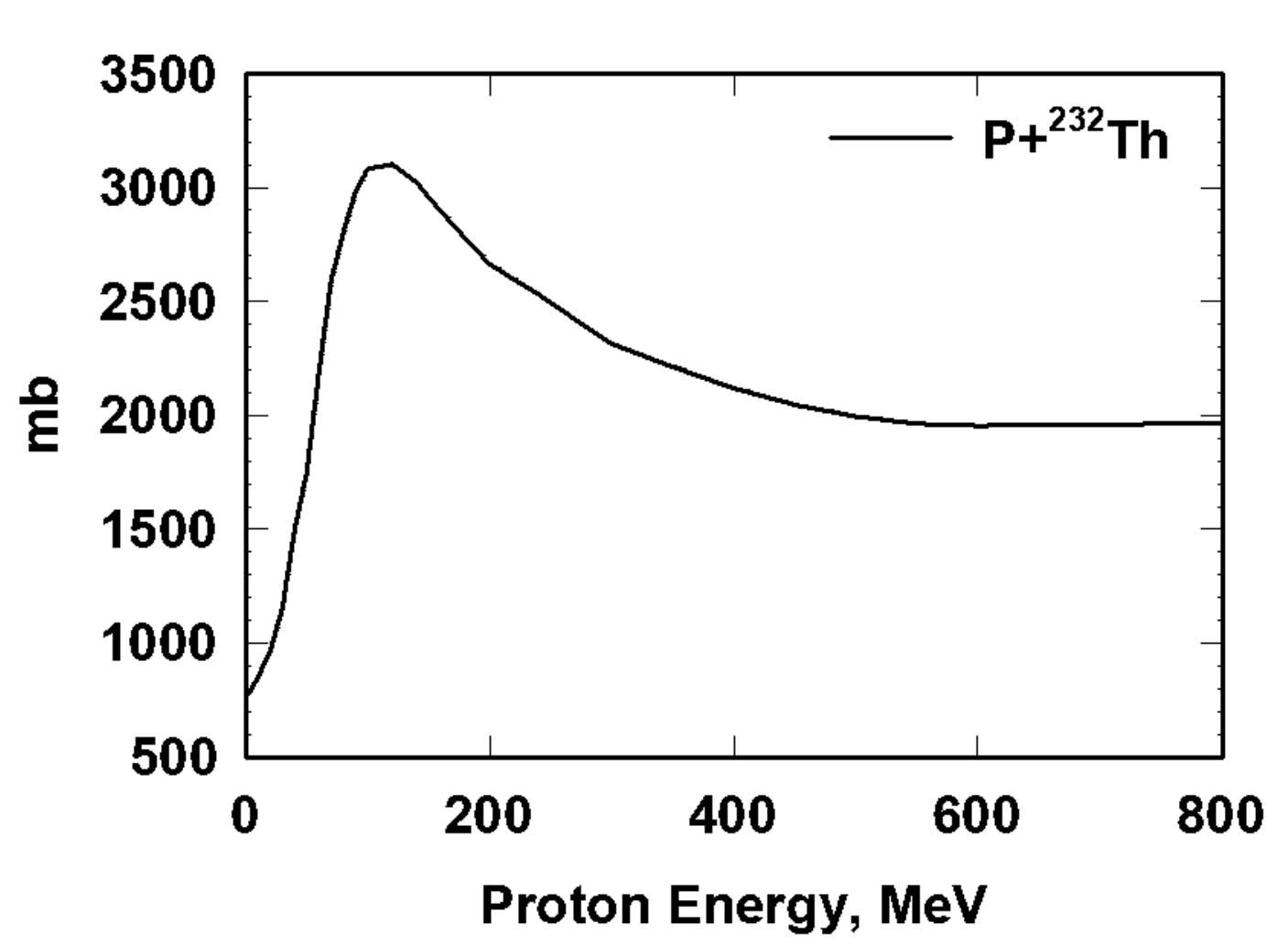}
\caption{Reaction cross-section for Proton+$^{232}$Th system using MONC} \label{pthcrs}
\end{figure}

In order to have quantitative estimate of $^{225}$Ac activity produced by proton+$^{232}$Th system, total reaction and individual channel cross-sections are calculated and compared with experimental data. The optimum energy from proton+$^{232}$Th reaction cross-section point of view is $\sim$ 150MeV which is shown in Fig.\ref{pthcrs}. The low energy proton accelerators with higher current are easy to build compared to high energy accelerators so we have studied the production of $^{225}$Ac by low energy proton beam.

\begin{figure}
\includegraphics[scale=0.55]{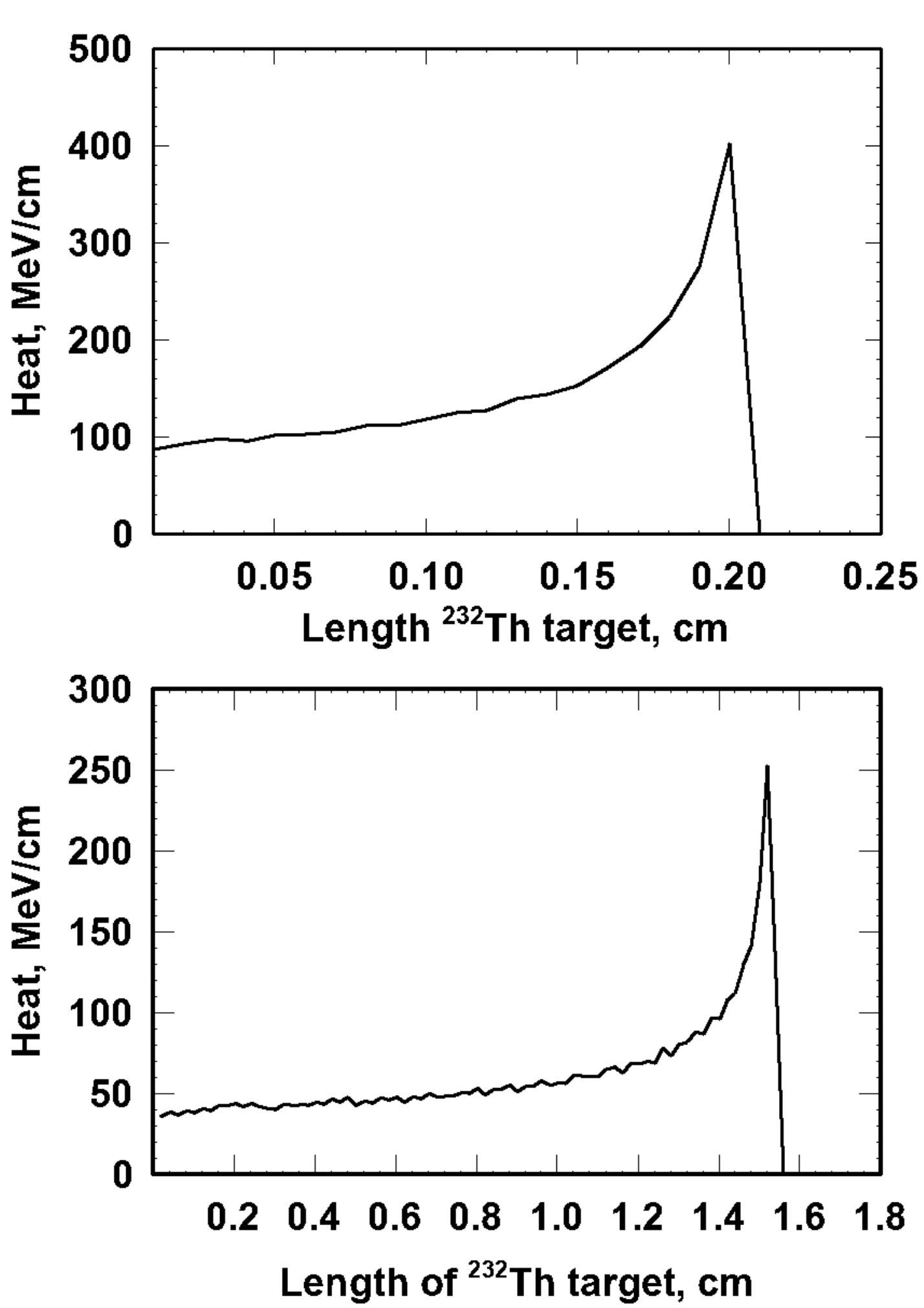}
\caption{Heat deposition and range of proton in $^{232}$Th target for 100MeV and 30MeV proton energies.} \label{heatth}
\end{figure}

\begin{figure}
\includegraphics[scale=0.37]{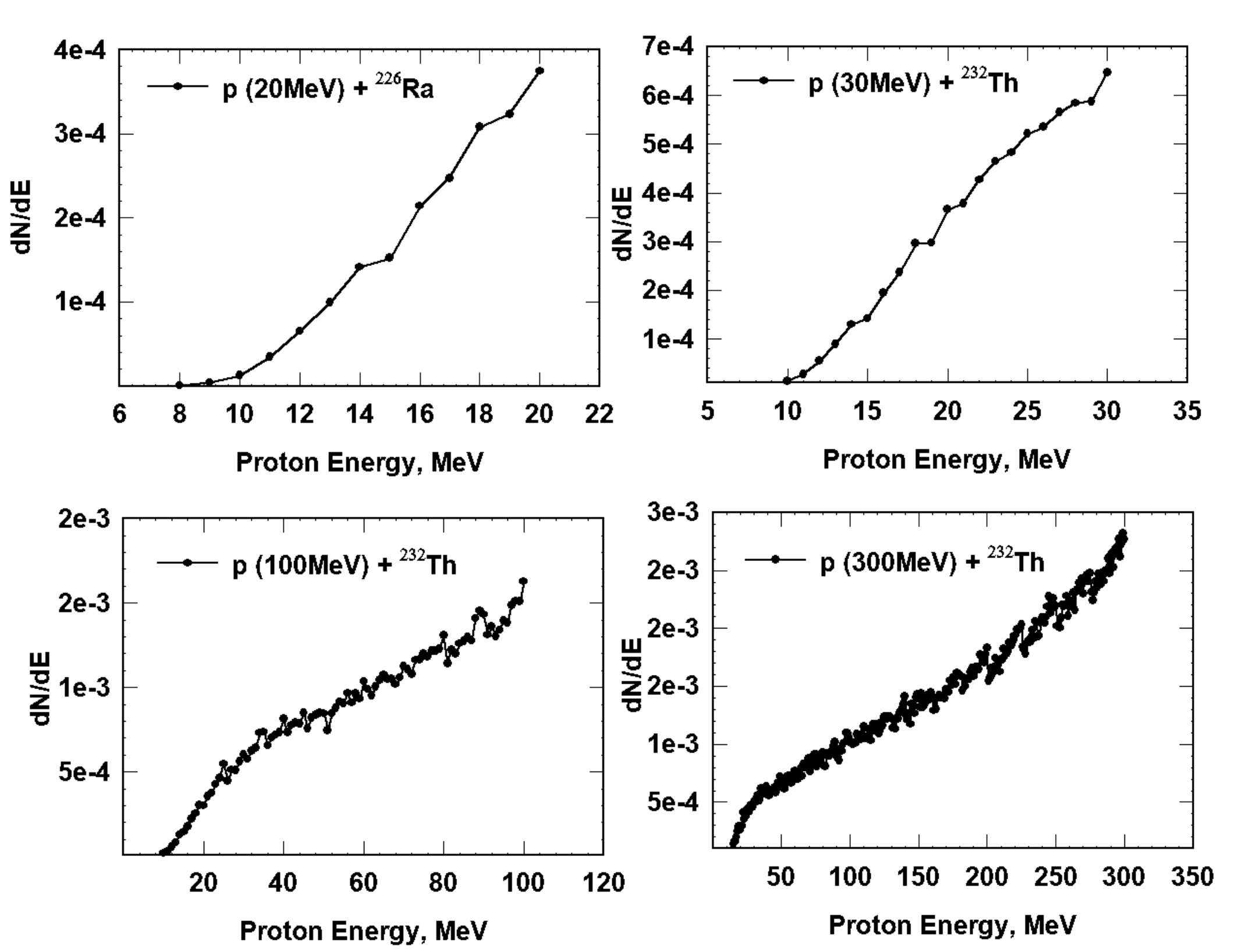}
\caption{Proton energy distribution after energy loss in thick $^{232}$Th target.} \label{heatth2}
\end{figure}

Economical production of $^{225}$Ac would require thick target irradiation leading to energy loss of the primary proton energy. The energy loss shows a range of $\sim$2mm and $\sim$16mm at 30MeV and 100MeV proton energies for Thorium target, as shown in Fig.\ref{heatth}. The energy of the primary proton also changes along its path in thick target due to continuous energy loss. The secondary protons undergoing nuclear reaction are shown in Fig. \ref{heatth2}.

More than 99$\%$ energy is deposited as heat at 30MeV which reduces to $\sim$96$\%$  at 100MeV for complete beam dump. The Temperature rise of Thorium metal of diameter 10cm and thickness 2mm after one hour irradiation of 30MeV proton beam with one $\mu$A current is calculated using the following rather simple Eq.\ref{temp} without considering any cooling.

\begin{equation}
 ms\Delta T = Q (Heat) \label{temp}
\end{equation}

Here, s=0.12J/gK is specific heat of Thorium metal, m is mass and Q is heat deposited by proton beam. The temperature comes out to be ~5000K which demands dedicated cooling of the target if it is used as beam dump. A small amount of Thorium as thin foil can be irradiated to avoid such large heating.

\begin{figure}
\includegraphics[scale=0.4]{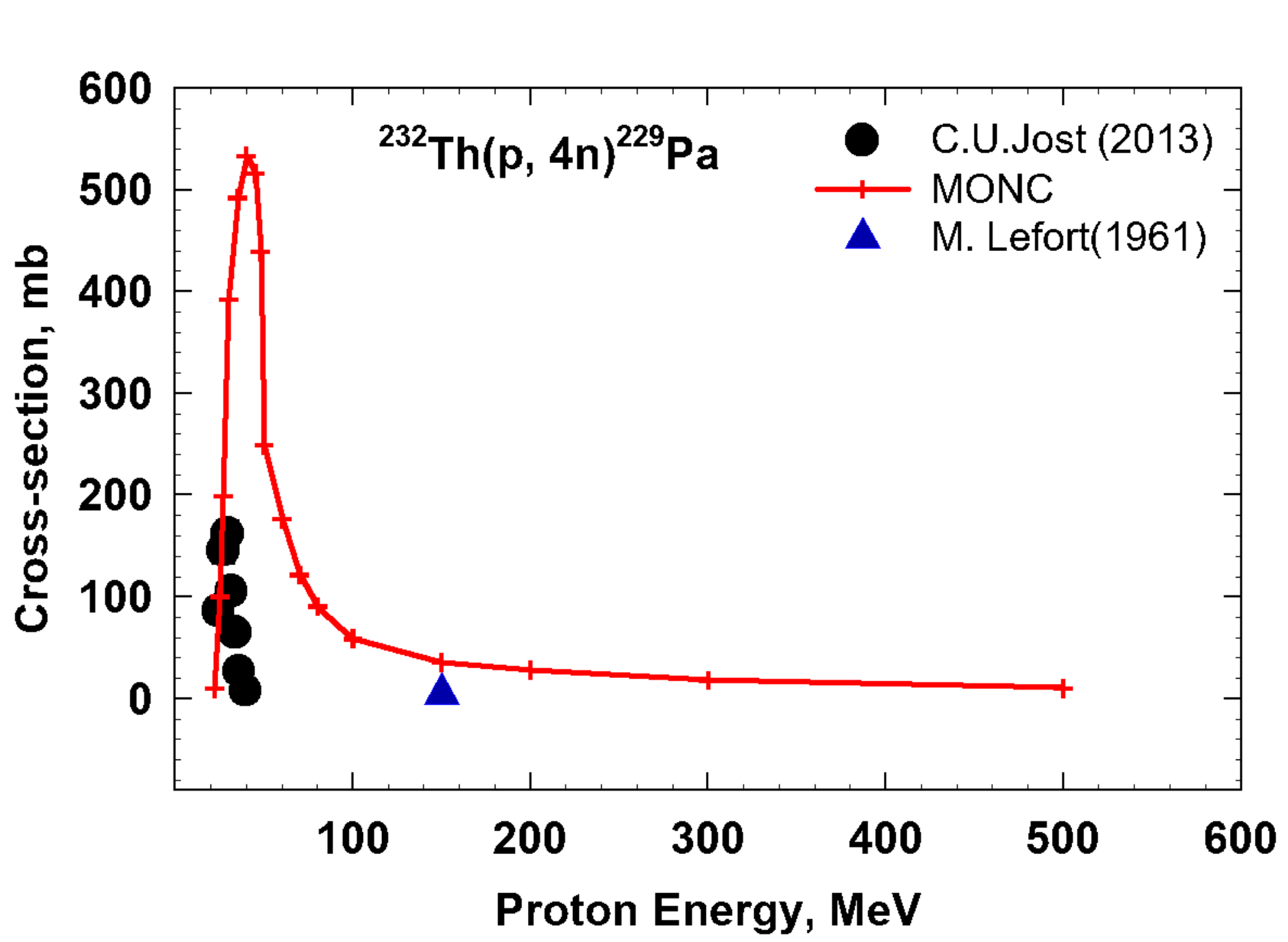}
\caption{Excitation function/production cross-section for the $^{232}$Th(p, X)$^{229}$Pa reaction.} \label{pthpa229}
\end{figure}

\begin{figure}
\includegraphics[scale=0.4]{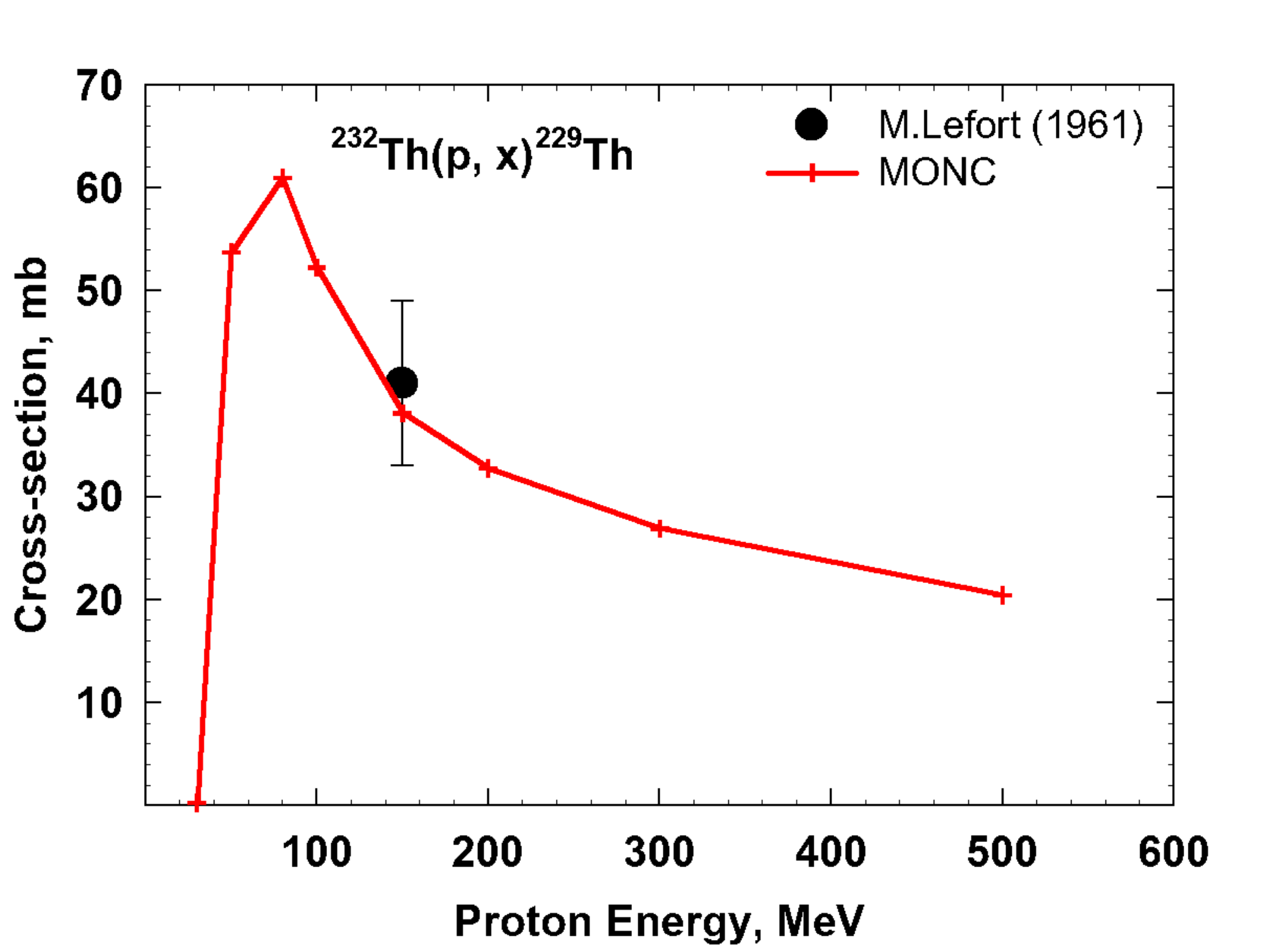}
\caption{Excitation function/production cross-section for the $^{232}$Th(p, X)$^{229}$Th reaction.} \label{pthth229}
\end{figure}

\begin{figure}
\includegraphics[scale=0.4]{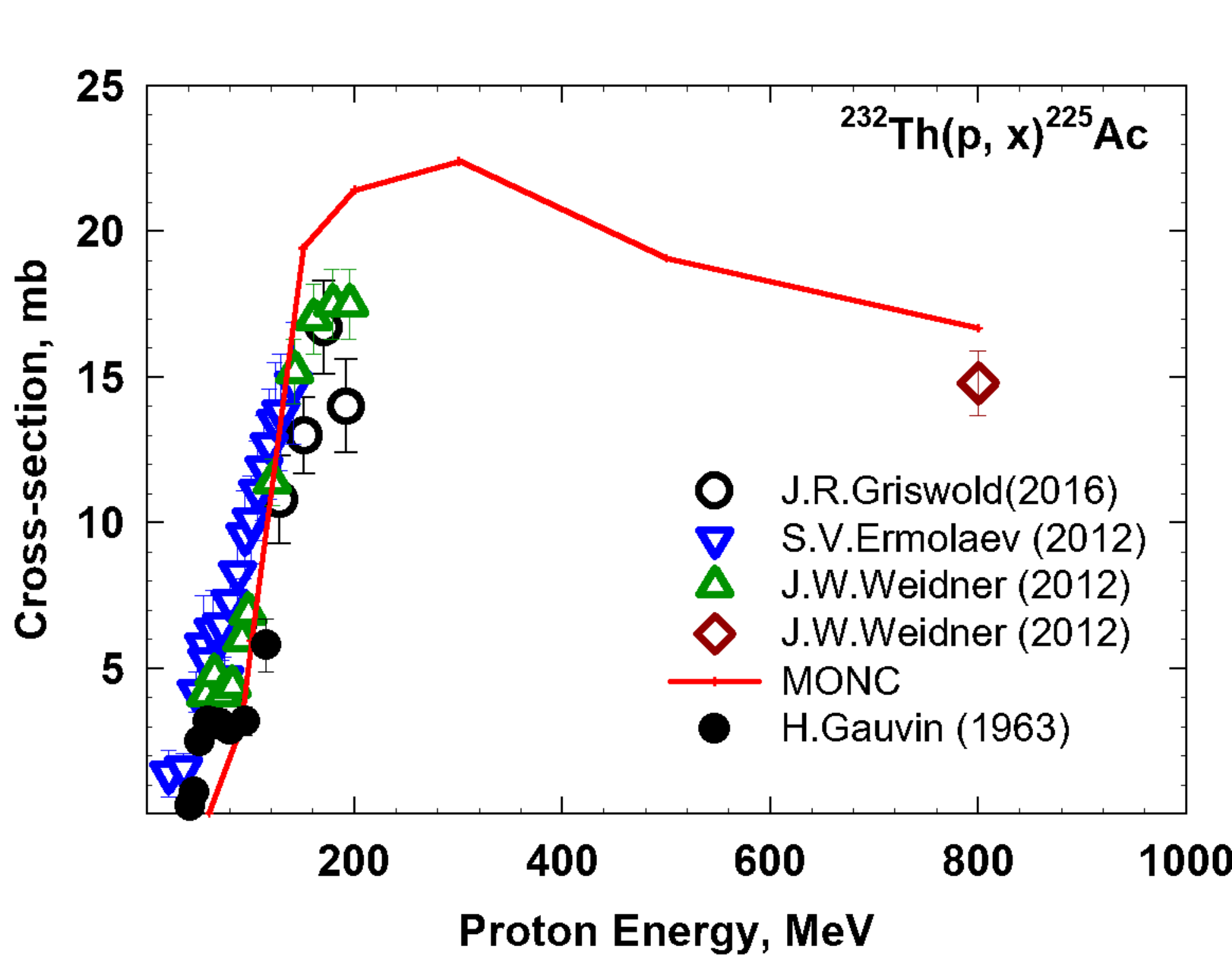}
\caption{Excitation function/production cross-section for the $^{232}$Th(p, X)$^{225}$Ac (cumulative yield) reaction.} \label{pthac225cm}
\end{figure}

Individual channel cross-section in MONC are based on the isotope production cross-section but not identification of channel in particular. The excitation function for the production of $^{229}$Pa, $^{229}$Th and $^{225}$Ac for p+$^{232}$Th system are compared with experimental data in Figs.\ref{pthpa229}, \ref{pthth229}, \ref{pthac225cm}, respectively. The cross-section data for $^{229}$Pa are taken from EXFOR data base maintained by IAEA \cite{Jost13,Lefort61}. The estimation from MONC for this isotope compared well at low energies of current interest but it overestimates between 50MeV to 100MeV range. Direct production of $^{229}$Th leading to $^{225}$Ac with long half life also has significant cross-section \cite{Lefort61} which is very well reproduced by MONC, although it has only single measured data point reported. Direct production of $^{225}$Ac has higher threshold and the measured cross-section \cite {Ermolaev12,Weidner2590,Mashnik2602,Griswold16} is also low. It should be noted that experimental data are derived from cumulative yield but MONC calculations are for independent yields. The measured data for independent yield are given by \cite{Gauvin} and depicted in the Fig.\ref{pthac225cm} by solid circles. It is to be noted that MONC compared well with these data points although the difference in cumulative and independent data is rather small.

\begin{figure}
\includegraphics[scale=0.4]{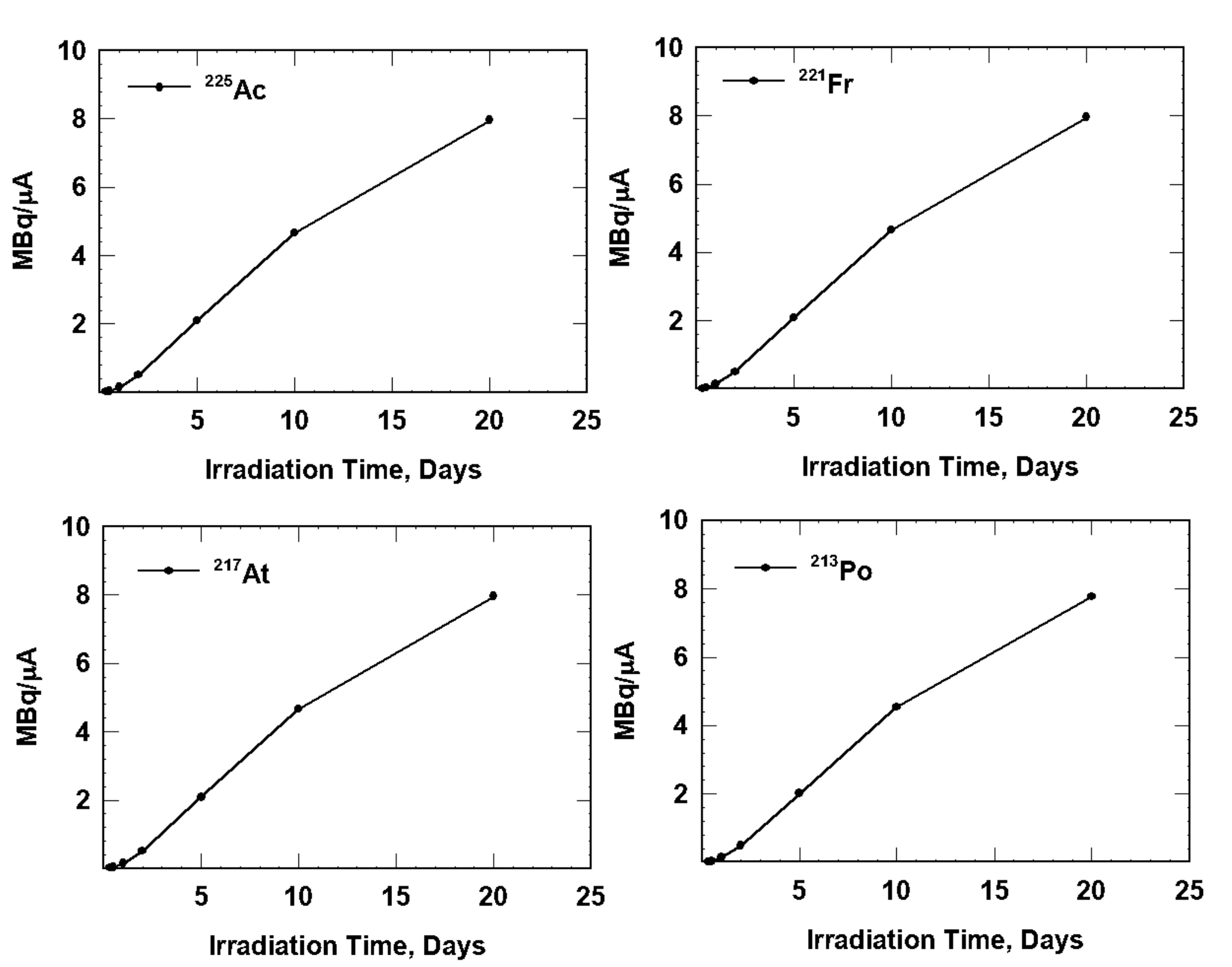}
\caption{Production of $^{225}$Ac and its $\alpha$-emitter decay products after different irradiation time using proton beam of 30MeV with 1$\mu$A current for p+$^{232}$Th system} \label{ac2251}. 
\end{figure}

After comparing the cross-section and heat deposition, produced activity for a thick target are estimated at different irradiation time periods. Simulation was performed for full beam deposition. Production of $^{225}$Ac in MBq per $\mu$A proton current of 30MeV energy and it's daughter products which are also $\alpha$-emitters are given in Fig.\ref{ac2251}. The combined activity for 10 days irradiation is $\sim$ 20MBq per $\mu$A which can be increased to 2GBq for 100 $\mu$A current. The main contributor of this production is $^{229}$Pa isotope as shown in Table \ref{tab:keffur}.

\begin{table}
\caption{\label{tab:keffur} Contribution of different decay channel in forming $^{225}$Ac at 30MeV proton energy for p+$^{232}$Th system.}
\begin{ruledtabular}
\begin{tabular}{ccc}
Target Thickness (mm)& Parent nucleus &$\%$ share\\\hline
 1.0    & $^{229}$Th    & 6.4452139E-08\\
 1.0    & $^{233}$Th    & 8.5865402E-15\\
 1.0    & $^{229}$Pa    & 0.999999936\\
 1.0    & $^{229}$Th    & 8.9715132E-14\\
\end{tabular}
\end{ruledtabular}
\end{table}

\begin{figure}
\mbox{\subfigure{\includegraphics[scale=0.5]{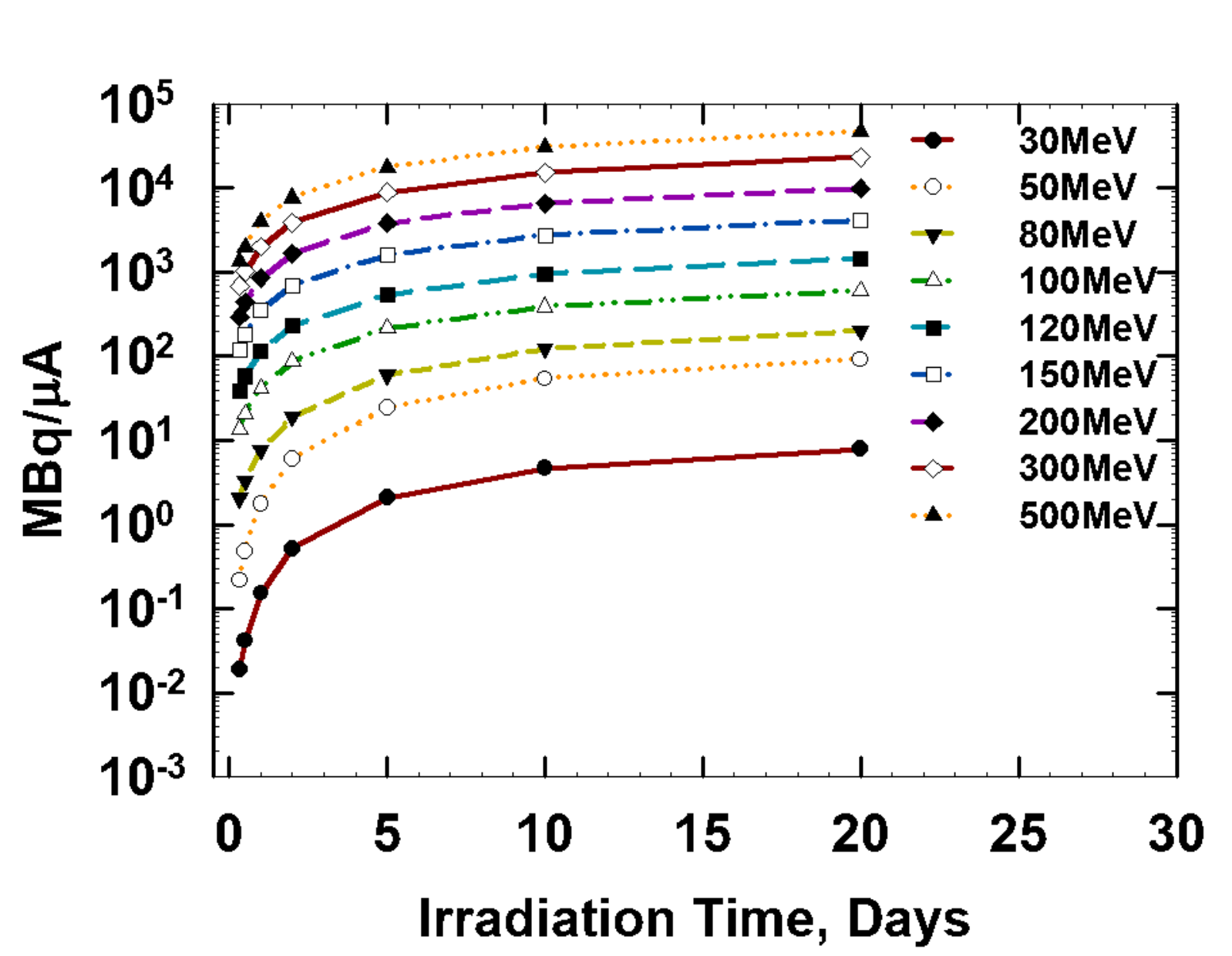}}}
\quad\subfigure{\includegraphics[scale=0.5]{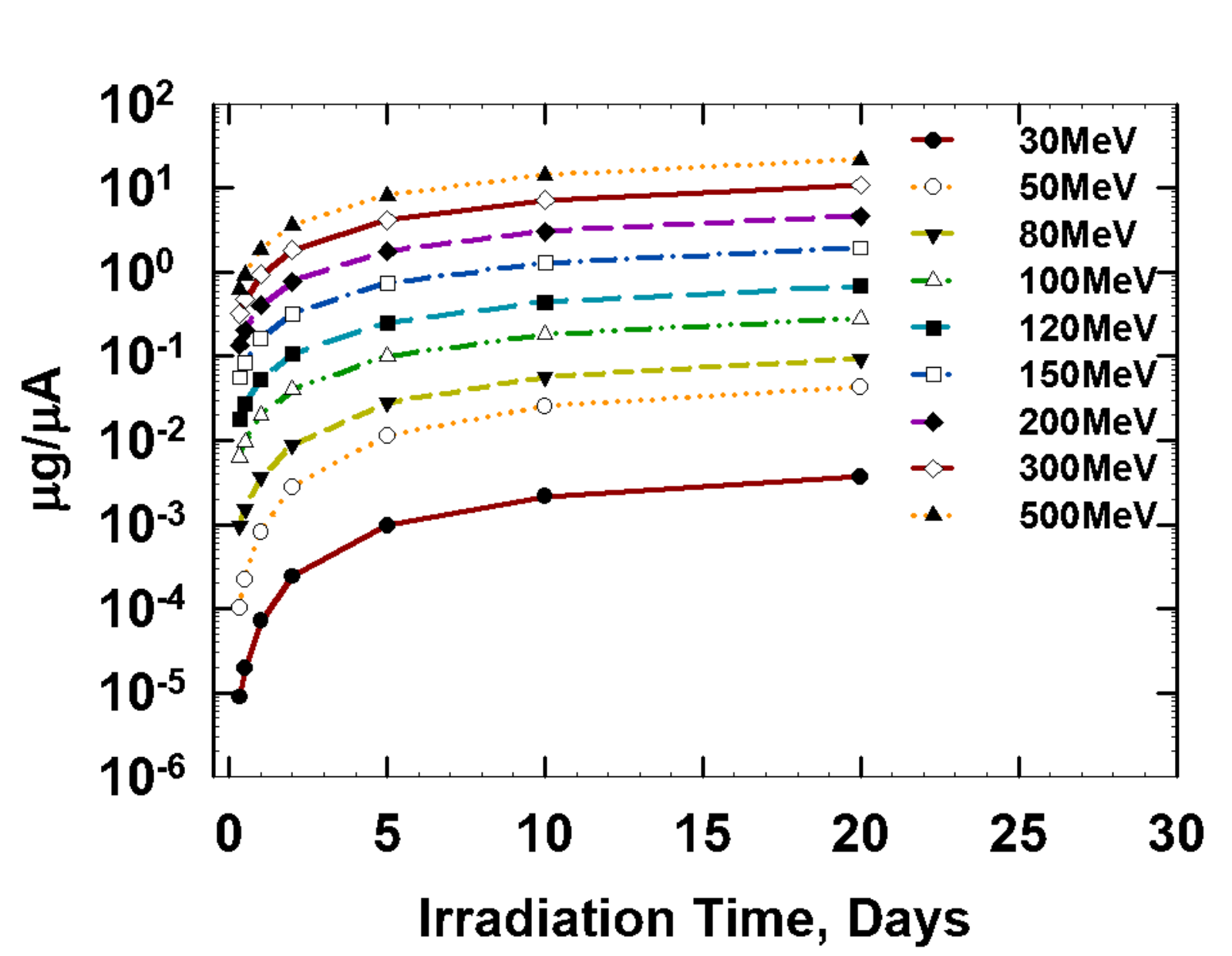}}
\caption{Production of $^{225}$Ac at different energies with 1$\mu$A current for p+$^{232}$Th system} \label{ac225_1}
\end{figure}

\begin{figure}
\includegraphics[scale=0.6]{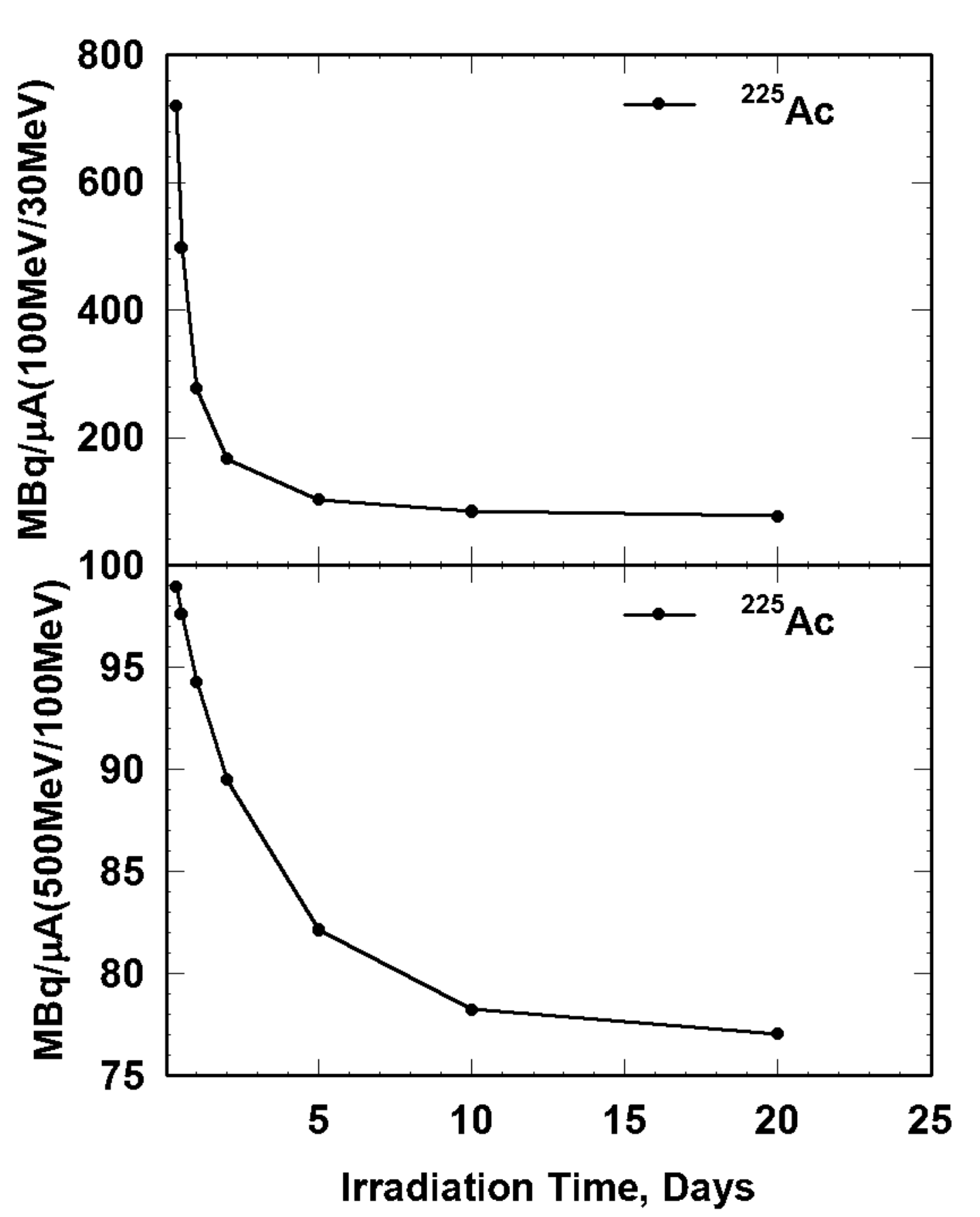}
\caption{Relative yield of $^{225}$Ac at 30MeV, 100MeV, and 500MeV for 1$\mu$A current for p+$^{232}$Th system.} \label{ac2252}
\end{figure}

\begin{figure}
 \includegraphics[scale=0.4]{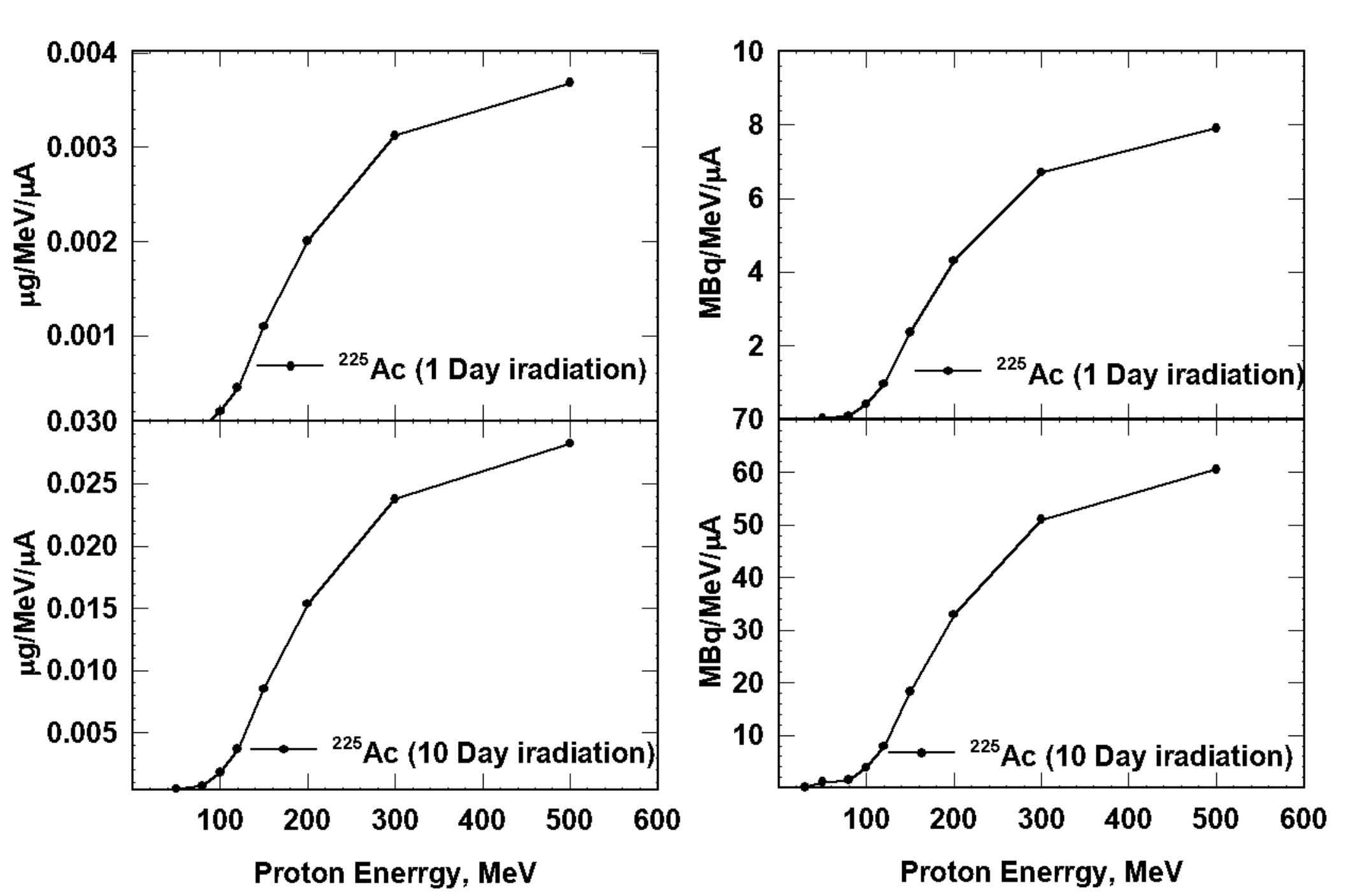}
 \caption{proton energy spent per micro gram production of $^{225}$Ac for one and 10 day accelerator operation with 1$\mu$A current for p+$^{232}$Th system.} \label{ac2253}
 \end{figure}

A comparative study for full proton beam dump in $^{232}$Th target shows significantly increased activity of $^{225}$Ac with energy as the cross-section for direct production of $^{225}$Ac increases with proton energy. The activity continuously increases with increase in energy as shown Fig.\ref{ac225_1}. Ratio of $^{225}$Ac activity at 100MeV and 30MeV shows a gain of $\sim$ 100 for complete dump of the beam. The gain factor for 300MeV compared to 100MeV is $\sim$ 80 for full beam dump in the target as shown in Fig.\ref{ac2252}. Although there is an increase in activity with proton energy increment but energy spent per MBq of $^{225}$Ac produced in this reaction as shown in Fig.\ref{ac2253}, shows that higher energies are not attractive as no significant gain in activity after 300MeV which may be termed as most optimum energy. 

\begin{figure}
\includegraphics[scale=0.4]{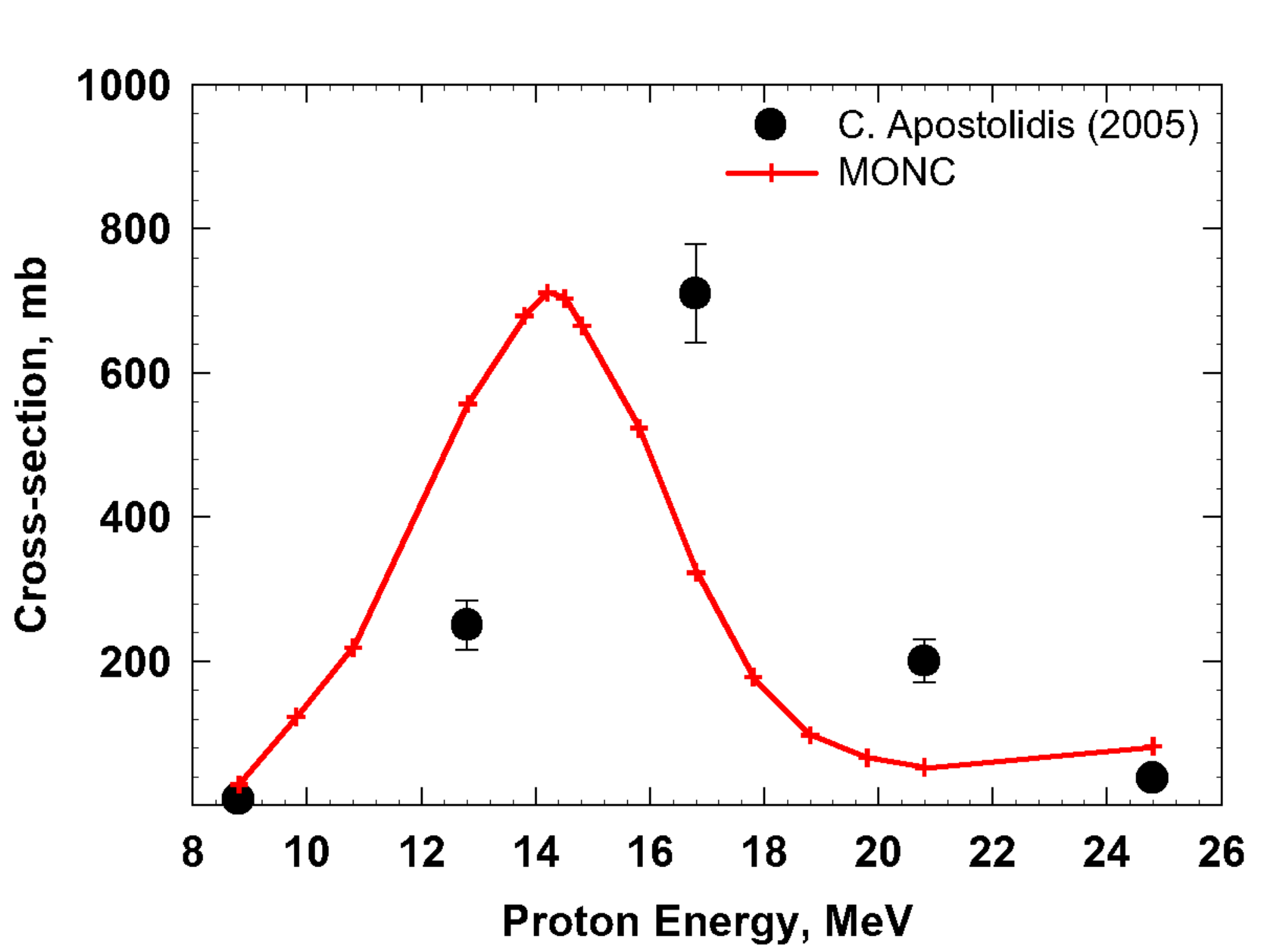}
\caption{calculated reaction $^{226}$Ra(p,2n)$^{225}$Ac cross-section using MONC} \label{pracrs}
\end{figure}

\begin{table}
\caption{\label{tab:keffur2} Production of $^{225}$Ac for 1$\mu$A proton beam current of different energies, 1mm target thickness and for different irradiation times.}
\begin{ruledtabular}
\begin{tabular}{ccccc}
Target &Thickness& Activity & Irradiation & Energy\\
       &(mm)& (MBq/$\mu$A) &  (Days)& (MeV)\\\hline
$^{232}$Th &1.0    &  5.2 & 10 &30\\
$^{232}$Th &1.0    &  9 & 20 &30\\
$^{232}$Th &1.0    &  11 & 30 &30\\
$^{232}$Th &1.0    &  83 & 10 &100\\
$^{226}$Ra &1.0    &  1643 & 10 &20\\
$^{226}$Ra &1.0    &  2465 & 20 &20\\
$^{226}$Ra &1.0    &  2876 & 30 &20\\
$^{226}$Ra &1.0    &  1709 & 10 &30\\

\end{tabular}
\end{ruledtabular}
\end{table}

\begin{figure}
\includegraphics[scale=0.4]{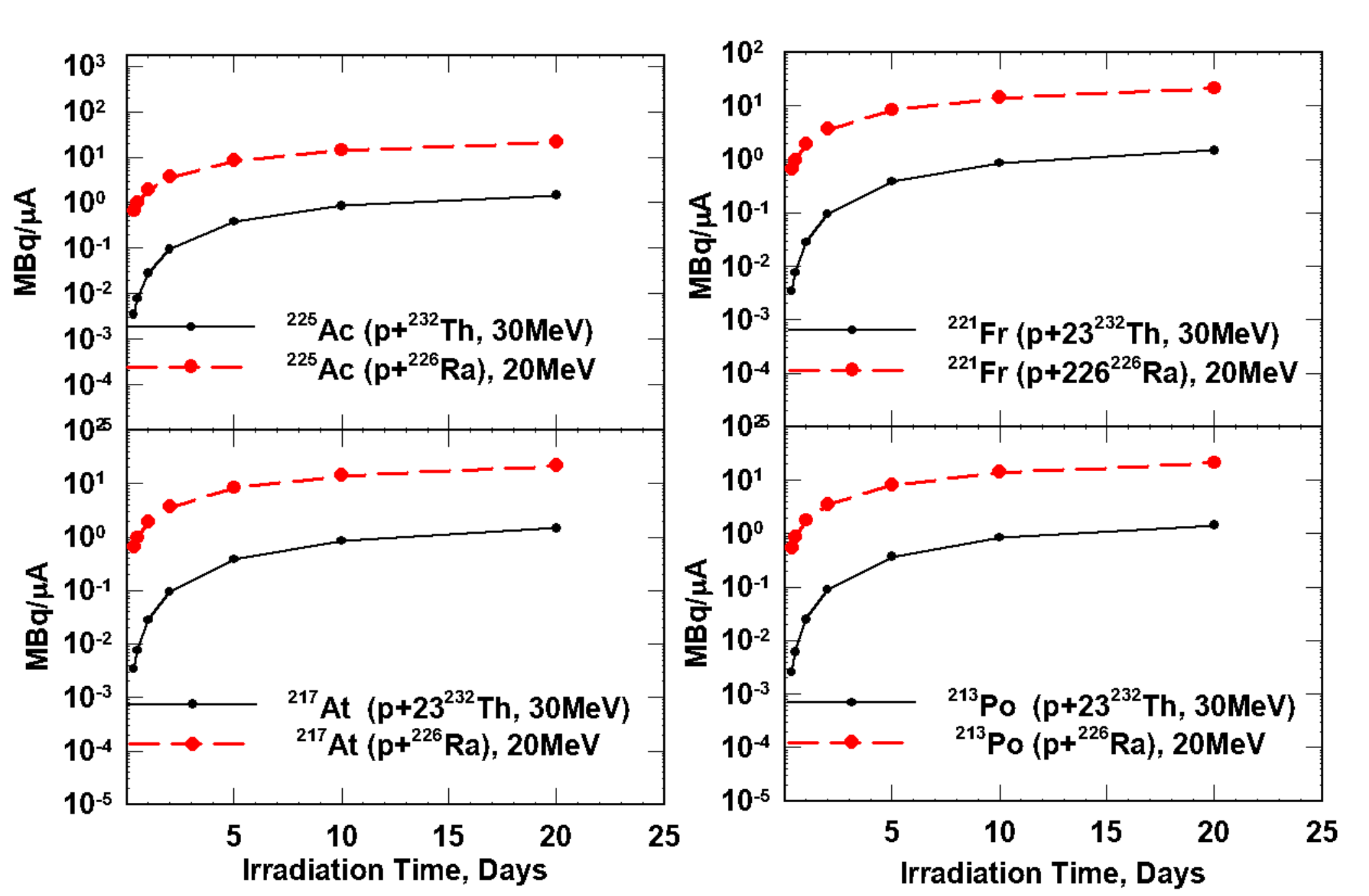}
\caption{Production of $^{225}$Ac and its $\alpha-emitter$ decay products after different irradiation time using p(30MeV) + $^{232}$Th and  p(20MeV) + $^{226}$Ra with 1$\mu$A current} \label{ac225thra}
\end{figure}
Other reaction channel with $^{226}$Ra as target material which gives rise to $^{225}$Ac at rather low energy is as follows
\begin{equation}
 p + ^{226}Ra \Rightarrow 2n + ^{225}Ac  (\Delta Q = -6.823 MeV) \label{eqra229}
\end{equation}
The production cross-sections/excitation function are compared with the experimental data \cite{Apostolidis05} for this reaction as given in Fig.\ref{pracrs}. The area under peak is in good agreement with the experimental data with slight shift in the peak position. The activity (MBq/$\mu$A) of $^{225}$Ac is compared in Table \ref{tab:keffur2} for 1mm thick $^{232}$Th and $^{226}$Ra targets at different beam energies and irradiation times. It is clear from Table \ref{tab:keffur2} that the $^{226}$Ra target can give much higher activity compared to $^{232}$Th target and it is also much cleaner from co-produced radio isotopes. The Activity of the four $\alpha$-emitting isotopes is compared as shown in Fig.\ref{ac225thra} where the combined activity may meet the demand of the world community.


\section{\label{sec:level4} Conclusion}
The cross-section of different channel reactions p+$^{232}$Th systems has been calculated with MONC code and compared with experimental data and the results are in good agreement. Further, production of $^{225}$Ac is analysed at different proton energies for $^{232}$Th target material for full beam dump and 1mm thick targets. There are many challenging tasks like removal of other Actinium isotopes and others which may be difficult to separate. The production of $^{225}$Ac activity is very high for $^{226}$Ra target and seems very promising from practical and economical use, whereas availability and handling of Radium-226 target might be a challenge. The produced activity of $^{225}$Ac in a week of irradiation through $^{226}$Ra target may be equal to present total annual production by milking $^{229}$Th. Thorium yields smaller $^{225}$Ac activity at low proton energies ($\sim$30MeV) but one can be benefited from higher current, stable proton accelerators.  Proton energy of 100MeV or above with hundreds of $\mu$A current can produce reasonable $^{225}$Ac activity for practical use. The contamination of other co-produced radio-isotopes is much less at low proton energies hence low energy high current proton accelerators might also be a viable alternative to generate Actinium-225 radio isotope using $^{232}$Th target.

\begin{acknowledgments}
The authors are immensely thankful to Dr. A. K. Mohanty (Director, BARC) to support this work.
\end{acknowledgments}

\begin{thebibliography}{27}%
\makeatletter
\providecommand \@ifxundefined [1]{%
 \@ifx{#1\undefined}
}%
\providecommand \@ifnum [1]{%
 \ifnum #1\expandafter \@firstoftwo
 \else \expandafter \@secondoftwo
 \fi
}%
\providecommand \@ifx [1]{%
 \ifx #1\expandafter \@firstoftwo
 \else \expandafter \@secondoftwo
 \fi
}%
\providecommand \natexlab [1]{#1}%
\providecommand \enquote  [1]{``#1''}%
\providecommand \bibnamefont  [1]{#1}%
\providecommand \bibfnamefont [1]{#1}%
\providecommand \citenamefont [1]{#1}%
\providecommand \href@noop [0]{\@secondoftwo}%
\providecommand \href [0]{\begingroup \@sanitize@url \@href}%
\providecommand \@href[1]{\@@startlink{#1}\@@href}%
\providecommand \@@href[1]{\endgroup#1\@@endlink}%
\providecommand \@sanitize@url [0]{\catcode `\\12\catcode `\$12\catcode
  `\&12\catcode `\#12\catcode `\^12\catcode `\_12\catcode `\%12\relax}%
\providecommand \@@startlink[1]{}%
\providecommand \@@endlink[0]{}%
\providecommand \url  [0]{\begingroup\@sanitize@url \@url }%
\providecommand \@url [1]{\endgroup\@href {#1}{\urlprefix }}%
\providecommand \urlprefix  [0]{URL }%
\providecommand \Eprint [0]{\href }%
\providecommand \doibase [0]{https://doi.org/}%
\providecommand \selectlanguage [0]{\@gobble}%
\providecommand \bibinfo  [0]{\@secondoftwo}%
\providecommand \bibfield  [0]{\@secondoftwo}%
\providecommand \translation [1]{[#1]}%
\providecommand \BibitemOpen [0]{}%
\providecommand \bibitemStop [0]{}%
\providecommand \bibitemNoStop [0]{.\EOS\space}%
\providecommand \EOS [0]{\spacefactor3000\relax}%
\providecommand \BibitemShut  [1]{\csname bibitem#1\endcsname}%
\let\auto@bib@innerbib\@empty
\bibitem [{\citenamefont {Humm}\ and\ \citenamefont {Cobb}(1990)}]{humm90}%
  \BibitemOpen
  \bibfield  {author} {\bibinfo {author} {\bibfnamefont {J.~L.}\ \bibnamefont
  {Humm}}\ and\ \bibinfo {author} {\bibfnamefont {L.~M.}\ \bibnamefont
  {Cobb}},\ }\href@noop {} {\bibfield  {journal} {\bibinfo  {journal} {J. Nucl.
  Med.}\ }\textbf {\bibinfo {volume} {31}},\ \bibinfo {pages} {75} (\bibinfo
  {year} {1990})}\BibitemShut {NoStop}%
\bibitem [{\citenamefont {Sgouros}\ \emph {et~al.}(2010)\citenamefont
  {Sgouros}, \citenamefont {Roeske}, \citenamefont {McDevitt}, \citenamefont
  {Palm}, \citenamefont {Allen}, \citenamefont {Fisher}, \citenamefont {Brill},
  \citenamefont {Song}, \citenamefont {Howell}, \citenamefont {Akabani},
  \citenamefont {SNM MIRD~Committee}, \citenamefont {Brill}, \citenamefont
  {Fisher}, \citenamefont {Howell}, \citenamefont {Meredith}, \citenamefont
  {Sgouros}, \citenamefont {Wessels},\ and\ \citenamefont
  {Zanzonico}}]{sgouros1}%
  \BibitemOpen
  \bibfield  {author} {\bibinfo {author} {\bibfnamefont {G.}~\bibnamefont
  {Sgouros}}, \bibinfo {author} {\bibfnamefont {J.}~\bibnamefont {Roeske}},
  \bibinfo {author} {\bibfnamefont {M.}~\bibnamefont {McDevitt}}, \bibinfo
  {author} {\bibfnamefont {S.}~\bibnamefont {Palm}}, \bibinfo {author}
  {\bibfnamefont {B.}~\bibnamefont {Allen}}, \bibinfo {author} {\bibfnamefont
  {D.}~\bibnamefont {Fisher}}, \bibinfo {author} {\bibfnamefont
  {A.}~\bibnamefont {Brill}}, \bibinfo {author} {\bibfnamefont
  {H.}~\bibnamefont {Song}}, \bibinfo {author} {\bibfnamefont {R.}~\bibnamefont
  {Howell}}, \bibinfo {author} {\bibfnamefont {G.}~\bibnamefont {Akabani}},
  \bibinfo {author} {\bibfnamefont {W.}~\bibnamefont {SNM MIRD~Committee},
  \bibfnamefont {Bolch}}, \bibinfo {author} {\bibfnamefont {A.}~\bibnamefont
  {Brill}}, \bibinfo {author} {\bibfnamefont {D.}~\bibnamefont {Fisher}},
  \bibinfo {author} {\bibfnamefont {R.}~\bibnamefont {Howell}}, \bibinfo
  {author} {\bibfnamefont {R.}~\bibnamefont {Meredith}}, \bibinfo {author}
  {\bibfnamefont {G.}~\bibnamefont {Sgouros}}, \bibinfo {author} {\bibfnamefont
  {B.}~\bibnamefont {Wessels}},\ and\ \bibinfo {author} {\bibfnamefont
  {P.}~\bibnamefont {Zanzonico}},\ }\href@noop {} {\bibfield  {journal}
  {\bibinfo  {journal} {J. Nucl. Med.}\ }\textbf {\bibinfo {volume} {51}},\
  \bibinfo {pages} {311} (\bibinfo {year} {2010})}\BibitemShut {NoStop}%
\bibitem [{\citenamefont {Essler}\ \emph {et~al.}(2012)\citenamefont {Essler},
  \citenamefont {Gärtner}, \citenamefont {Neff}, \citenamefont {Blechert},
  \citenamefont {Senekowitsch-Schmidtke}, \citenamefont {Bruchertseifer},
  \citenamefont {Morgenstern},\ and\ \citenamefont {Seidl}}]{essler12}%
  \BibitemOpen
  \bibfield  {author} {\bibinfo {author} {\bibfnamefont {M.}~\bibnamefont
  {Essler}}, \bibinfo {author} {\bibfnamefont {F.}~\bibnamefont {Gärtner}},
  \bibinfo {author} {\bibfnamefont {F.}~\bibnamefont {Neff}}, \bibinfo {author}
  {\bibfnamefont {B.}~\bibnamefont {Blechert}}, \bibinfo {author}
  {\bibfnamefont {R.}~\bibnamefont {Senekowitsch-Schmidtke}}, \bibinfo {author}
  {\bibfnamefont {F.}~\bibnamefont {Bruchertseifer}}, \bibinfo {author}
  {\bibfnamefont {A.}~\bibnamefont {Morgenstern}},\ and\ \bibinfo {author}
  {\bibfnamefont {C.}~\bibnamefont {Seidl}},\ }\href@noop {} {\bibfield
  {journal} {\bibinfo  {journal} {Eur. J. Nucl. Med. Mol. Imaging}\ }\textbf
  {\bibinfo {volume} {39}},\ \bibinfo {pages} {602–612} (\bibinfo {year}
  {2012})}\BibitemShut {NoStop}%
\bibitem [{\citenamefont {Brechbiel}(2007)}]{Brechbiel07}%
  \BibitemOpen
  \bibfield  {author} {\bibinfo {author} {\bibfnamefont {M.}~\bibnamefont
  {Brechbiel}},\ }\href@noop {} {\bibfield  {journal} {\bibinfo  {journal}
  {Dalton Trans.}\ ,\ \bibinfo {pages} {4918–4928}} (\bibinfo {year}
  {2007})}\BibitemShut {NoStop}%
\bibitem [{\citenamefont {Azure}\ \emph {et~al.}(1994)\citenamefont {Azure},
  \citenamefont {Archer}, \citenamefont {Sastry}, \citenamefont {Rao},\ and\
  \citenamefont {Howell}}]{Azure94}%
  \BibitemOpen
  \bibfield  {author} {\bibinfo {author} {\bibfnamefont {M.~T.}\ \bibnamefont
  {Azure}}, \bibinfo {author} {\bibfnamefont {R.~D.}\ \bibnamefont {Archer}},
  \bibinfo {author} {\bibfnamefont {K.~S.~R.}\ \bibnamefont {Sastry}}, \bibinfo
  {author} {\bibfnamefont {D.~V.}\ \bibnamefont {Rao}},\ and\ \bibinfo {author}
  {\bibfnamefont {R.~W.}\ \bibnamefont {Howell}},\ }\href@noop {} {\bibfield
  {journal} {\bibinfo  {journal} {Radiat. Res.}\ }\textbf {\bibinfo {volume}
  {140}},\ \bibinfo {pages} {276} (\bibinfo {year} {1994})}\BibitemShut
  {NoStop}%
\bibitem [{\citenamefont {McDevitt}(2001)}]{McDevitt01}%
  \BibitemOpen
  \bibfield  {author} {\bibinfo {author} {\bibfnamefont {M.~D. L. L. S. J. B.
  P. F. R. W. K. P. V. C. M. M. M. e.~a.}\ \bibnamefont {McDevitt},
  \bibfnamefont {M.R.}},\ }\href@noop {} {\bibfield  {journal} {\bibinfo
  {journal} {Science}\ }\textbf {\bibinfo {volume} {294}},\ \bibinfo {pages}
  {1537–1540} (\bibinfo {year} {2001})}\BibitemShut {NoStop}%
\bibitem [{iae(2013)}]{iaea2013}%
  \BibitemOpen
  \href@noop {} {\emph {\bibinfo {title} {IAEA Technical Meeting on Alpha
  emitting radionuclides and radiopharmaceuticals for therapy}}} (\bibinfo
  {year} {2013})\BibitemShut {NoStop}%
\bibitem [{\citenamefont {Sinha}\ and\ \citenamefont
  {Kakodkar}(2006)}]{shina06}%
  \BibitemOpen
  \bibfield  {author} {\bibinfo {author} {\bibfnamefont {R.}~\bibnamefont
  {Sinha}}\ and\ \bibinfo {author} {\bibfnamefont {A.}~\bibnamefont
  {Kakodkar}},\ }\href@noop {} {\bibfield  {journal} {\bibinfo  {journal}
  {Nucl. Eng. and Des.}\ }\textbf {\bibinfo {volume} {236}},\ \bibinfo {pages}
  {683} (\bibinfo {year} {2006})}\BibitemShut {NoStop}%
\bibitem [{\citenamefont {Jost}\ \emph
  {et~al.}(2013{\natexlab{a}})\citenamefont {Jost}, \citenamefont {Griswold},
  \citenamefont {Bruffey}, \citenamefont {Mirzadeh}, \citenamefont
  {S.Stracener},\ and\ \citenamefont {Williams}}]{jost2013}%
  \BibitemOpen
  \bibfield  {author} {\bibinfo {author} {\bibfnamefont {C.}~\bibnamefont
  {Jost}}, \bibinfo {author} {\bibfnamefont {J.}~\bibnamefont {Griswold}},
  \bibinfo {author} {\bibfnamefont {S.}~\bibnamefont {Bruffey}}, \bibinfo
  {author} {\bibnamefont {Mirzadeh}}, \bibinfo {author} {\bibfnamefont
  {D.}~\bibnamefont {S.Stracener}},\ and\ \bibinfo {author} {\bibfnamefont
  {C.}~\bibnamefont {Williams}},\ }\href@noop {} {\bibfield  {journal}
  {\bibinfo  {journal} {AIP Conference Proceedings: International Conference on
  Application of Accelerators in Research and Industry}\ }\textbf {\bibinfo
  {volume} {1525}} (\bibinfo {year} {2013}{\natexlab{a}})}\BibitemShut
  {NoStop}%
\bibitem [{\citenamefont {Apostolidis}\ \emph {et~al.}(2005)\citenamefont
  {Apostolidis}, \citenamefont {Molinet}, \citenamefont {McGinley},
  \citenamefont {Abbas}, \citenamefont {Möllenbeck},\ and\ \citenamefont
  {Morgenstern}}]{Apostolidis05}%
  \BibitemOpen
  \bibfield  {author} {\bibinfo {author} {\bibfnamefont {C.}~\bibnamefont
  {Apostolidis}}, \bibinfo {author} {\bibfnamefont {R.}~\bibnamefont
  {Molinet}}, \bibinfo {author} {\bibfnamefont {J.}~\bibnamefont {McGinley}},
  \bibinfo {author} {\bibfnamefont {K.}~\bibnamefont {Abbas}}, \bibinfo
  {author} {\bibfnamefont {J.}~\bibnamefont {Möllenbeck}},\ and\ \bibinfo
  {author} {\bibfnamefont {A.}~\bibnamefont {Morgenstern}},\ }\href@noop {}
  {\bibfield  {journal} {\bibinfo  {journal} {Appl. Radiat. Isot.}\ }\textbf
  {\bibinfo {volume} {62}},\ \bibinfo {pages} {383–387} (\bibinfo {year}
  {2005})}\BibitemShut {NoStop}%
\bibitem [{\citenamefont {Melville}\ \emph {et~al.}(2007)\citenamefont
  {Melville}, \citenamefont {Meriarty}, \citenamefont {Metcalfe}, \citenamefont
  {Knittel},\ and\ \citenamefont {Allen}}]{Melville07}%
  \BibitemOpen
  \bibfield  {author} {\bibinfo {author} {\bibfnamefont {G.}~\bibnamefont
  {Melville}}, \bibinfo {author} {\bibfnamefont {H.}~\bibnamefont {Meriarty}},
  \bibinfo {author} {\bibfnamefont {P.}~\bibnamefont {Metcalfe}}, \bibinfo
  {author} {\bibfnamefont {T.}~\bibnamefont {Knittel}},\ and\ \bibinfo {author}
  {\bibfnamefont {B.}~\bibnamefont {Allen}},\ }\href@noop {} {\bibfield
  {journal} {\bibinfo  {journal} {Appl. Radiat. Isot.}\ }\textbf {\bibinfo
  {volume} {65}},\ \bibinfo {pages} {1014–1022} (\bibinfo {year}
  {2007})}\BibitemShut {NoStop}%
\bibitem [{\citenamefont {Kumawat}\ and\ \citenamefont
  {Barashenkov}(2005)}]{hkumawat05}%
  \BibitemOpen
  \bibfield  {author} {\bibinfo {author} {\bibfnamefont {H.}~\bibnamefont
  {Kumawat}}\ and\ \bibinfo {author} {\bibfnamefont {V.~S.}\ \bibnamefont
  {Barashenkov}},\ }\href@noop {} {\bibfield  {journal} {\bibinfo  {journal}
  {Euro. Phys. J.}\ }\textbf {\bibinfo {volume} {A26}},\ \bibinfo {pages} {61}
  (\bibinfo {year} {2005})}\BibitemShut {NoStop}%
\bibitem [{\citenamefont {Kumawat}(2004)}]{hkumawat04}%
  \BibitemOpen
  \bibfield  {author} {\bibinfo {author} {\bibfnamefont {H.}~\bibnamefont
  {Kumawat}},\ }\href@noop {} {\bibfield  {journal} {\bibinfo  {journal}
  {Development of Monte-Carlo Complex Program CASCADE and its Applications to
  Mathematical Modelling of Transport of Particles in Many Component Systems,
  PhD Thesis, JINR, Dubna and JINR preprint}\ }\textbf {\bibinfo {volume}
  {E11-2004-166}} (\bibinfo {year} {2004})}\BibitemShut {NoStop}%
\bibitem [{\citenamefont {Kumawat}\ \emph {et~al.}(2016)\citenamefont
  {Kumawat}, \citenamefont {Saxena},\ and\ \citenamefont {Carminiti}}]{hkroot}%
  \BibitemOpen
  \bibfield  {author} {\bibinfo {author} {\bibfnamefont {H.}~\bibnamefont
  {Kumawat}}, \bibinfo {author} {\bibfnamefont {A.}~\bibnamefont {Saxena}},\
  and\ \bibinfo {author} {\bibfnamefont {F.}~\bibnamefont {Carminiti}},\
  }\href@noop {} {\bibfield  {journal} {\bibinfo  {journal} {BARC External
  Report}\ }\textbf {\bibinfo {volume} {BARC/2016/E/017}} (\bibinfo {year}
  {2016})}\BibitemShut {NoStop}%
\bibitem [{\citenamefont {Kumawat}\ and\ \citenamefont
  {Venkata}(2013)}]{hkvenkata13}%
  \BibitemOpen
  \bibfield  {author} {\bibinfo {author} {\bibfnamefont {H.}~\bibnamefont
  {Kumawat}}\ and\ \bibinfo {author} {\bibfnamefont {P.~P.~K.}\ \bibnamefont
  {Venkata}},\ }\href@noop {} {\bibfield  {journal} {\bibinfo  {journal} {BARC
  NEWSLETTER}\ }\textbf {\bibinfo {volume} {332}} (\bibinfo {year}
  {2013})}\BibitemShut {NoStop}%
\bibitem [{\citenamefont {Barashenkov}\ and\ \citenamefont
  {Toneev}(1972)}]{baras1}%
  \BibitemOpen
  \bibfield  {author} {\bibinfo {author} {\bibfnamefont {V.}~\bibnamefont
  {Barashenkov}}\ and\ \bibinfo {author} {\bibfnamefont {V.}~\bibnamefont
  {Toneev}},\ }\href@noop {} {\bibfield  {journal} {\bibinfo  {journal}
  {Atomizdat, Moscow}\ } (\bibinfo {year} {1972})}\BibitemShut {NoStop}%
\bibitem [{\citenamefont {Mashnik}\ and\ \citenamefont
  {Toneev}(1974)}]{mashnik}%
  \BibitemOpen
  \bibfield  {author} {\bibinfo {author} {\bibfnamefont {S.}~\bibnamefont
  {Mashnik}}\ and\ \bibinfo {author} {\bibfnamefont {V.}~\bibnamefont
  {Toneev}},\ }\href@noop {} {\bibfield  {journal} {\bibinfo  {journal} {JINR,
  Dubna.}\ }\textbf {\bibinfo {volume} {P4-9417}} (\bibinfo {year}
  {1974})}\BibitemShut {NoStop}%
\bibitem [{\citenamefont {Kumawat}(2010)}]{hkumawat10}%
  \BibitemOpen
  \bibfield  {author} {\bibinfo {author} {\bibfnamefont {H.}~\bibnamefont
  {Kumawat}},\ }\href {http://www-nds.iaea.org/spallations/spal_mdl.html}
  {\emph {\bibinfo {title} {IAEA Benchmark of spallation models}}} (\bibinfo
  {year} {March 2010})\BibitemShut {NoStop}%
\bibitem [{\citenamefont {Kumawat\textit{ et al.}}(2008)}]{hkheat08}%
  \BibitemOpen
  \bibfield  {author} {\bibinfo {author} {\bibfnamefont {H.}~\bibnamefont
  {Kumawat\textit{ et al.}}},\ }\href@noop {} {\bibfield  {journal} {\bibinfo
  {journal} {Nucl. Instr. Meth. Phys. Res.}\ }\textbf {\bibinfo {volume}
  {B266}},\ \bibinfo {pages} {604} (\bibinfo {year} {2008})}\BibitemShut
  {NoStop}%
\bibitem [{\citenamefont {Kumawat}\ \emph {et~al.}(2009)\citenamefont
  {Kumawat}, \citenamefont {Srinivasan},\ and\ \citenamefont
  {Kumar}}]{hkumawat09}%
  \BibitemOpen
  \bibfield  {author} {\bibinfo {author} {\bibfnamefont {H.}~\bibnamefont
  {Kumawat}}, \bibinfo {author} {\bibfnamefont {P.}~\bibnamefont
  {Srinivasan}},\ and\ \bibinfo {author} {\bibfnamefont {V.}~\bibnamefont
  {Kumar}},\ }\href@noop {} {\bibfield  {journal} {\bibinfo  {journal} {Pramana
  J. Phys.}\ }\textbf {\bibinfo {volume} {72}},\ \bibinfo {pages} {601}
  (\bibinfo {year} {2009})}\BibitemShut {NoStop}%
\bibitem [{\citenamefont {Cetnar}(2006)}]{bateman06}%
  \BibitemOpen
  \bibfield  {author} {\bibinfo {author} {\bibfnamefont {J.}~\bibnamefont
  {Cetnar}},\ }\href@noop {} {\bibfield  {journal} {\bibinfo  {journal} {Ann.
  Nucl. Ene.}\ }\textbf {\bibinfo {volume} {33}},\ \bibinfo {pages} {640}
  (\bibinfo {year} {2006})}\BibitemShut {NoStop}%
\bibitem [{\citenamefont {Jost}\ \emph
  {et~al.}(2013{\natexlab{b}})\citenamefont {Jost}, \citenamefont {Griswold},
  \citenamefont {Bruffey}, \citenamefont {Mirzadeh}, \citenamefont
  {Stracener},\ and\ \citenamefont {Williams}}]{Jost13}%
  \BibitemOpen
  \bibfield  {author} {\bibinfo {author} {\bibfnamefont {C.~U.}\ \bibnamefont
  {Jost}}, \bibinfo {author} {\bibfnamefont {J.~R.}\ \bibnamefont {Griswold}},
  \bibinfo {author} {\bibfnamefont {S.~H.}\ \bibnamefont {Bruffey}}, \bibinfo
  {author} {\bibfnamefont {S.}~\bibnamefont {Mirzadeh}}, \bibinfo {author}
  {\bibfnamefont {D.~W.}\ \bibnamefont {Stracener}},\ and\ \bibinfo {author}
  {\bibfnamefont {C.~L.}\ \bibnamefont {Williams}},\ }\href@noop {} {}\bibinfo
  {type} {Tech. Rep.}\ \bibinfo {number} {1525}\ (\bibinfo {year} {2013})\
  \bibinfo {note} {conf.proceedings by Am.Inst.of Phys.}\BibitemShut {Stop}%
\bibitem [{\citenamefont {Lefort}\ \emph {et~al.}(1961)\citenamefont {Lefort},
  \citenamefont {Simonoff},\ and\ \citenamefont {Tarrago}}]{Lefort61}%
  \BibitemOpen
  \bibfield  {author} {\bibinfo {author} {\bibfnamefont {M.}~\bibnamefont
  {Lefort}}, \bibinfo {author} {\bibfnamefont {G.~N.}\ \bibnamefont
  {Simonoff}},\ and\ \bibinfo {author} {\bibfnamefont {X.}~\bibnamefont
  {Tarrago}},\ }\href@noop {} {\bibfield  {journal} {\bibinfo  {journal}
  {Nuclear Physics}\ }\textbf {\bibinfo {volume} {25}},\ \bibinfo {pages} {216}
  (\bibinfo {year} {1961})}\BibitemShut {NoStop}%
\bibitem [{\citenamefont {Ermolaev}\ \emph {et~al.}(2012)\citenamefont
  {Ermolaev}, \citenamefont {Zhuikov}, \citenamefont {Kokhanyuk}, \citenamefont
  {Matushko}, \citenamefont {Kalmykov}, \citenamefont {Aliev}, \citenamefont
  {Tananaev},\ and\ \citenamefont {Myasoedov}}]{Ermolaev12}%
  \BibitemOpen
  \bibfield  {author} {\bibinfo {author} {\bibfnamefont {S.~V.}\ \bibnamefont
  {Ermolaev}}, \bibinfo {author} {\bibfnamefont {B.~L.}\ \bibnamefont
  {Zhuikov}}, \bibinfo {author} {\bibfnamefont {V.~M.}\ \bibnamefont
  {Kokhanyuk}}, \bibinfo {author} {\bibfnamefont {V.~L.}\ \bibnamefont
  {Matushko}}, \bibinfo {author} {\bibfnamefont {S.~N.}\ \bibnamefont
  {Kalmykov}}, \bibinfo {author} {\bibfnamefont {R.~A.}\ \bibnamefont {Aliev}},
  \bibinfo {author} {\bibfnamefont {I.~G.}\ \bibnamefont {Tananaev}},\ and\
  \bibinfo {author} {\bibfnamefont {B.~F.}\ \bibnamefont {Myasoedov}},\
  }\href@noop {} {\bibfield  {journal} {\bibinfo  {journal} {Radiochimica
  Acta}\ }\textbf {\bibinfo {volume} {100}},\ \bibinfo {pages} {223} (\bibinfo
  {year} {2012})}\BibitemShut {NoStop}%
\bibitem [{\citenamefont {Weidner}\ \emph {et~al.}(2012)\citenamefont
  {Weidner}, \citenamefont {Mashnik}, \citenamefont {John}, \citenamefont
  {Hemez}, \citenamefont {Ballard}, \citenamefont {Bach}, \citenamefont
  {Birnbaum}, \citenamefont {Bitteker}, \citenamefont {Couture}, \citenamefont
  {Dry}, \citenamefont {Fassbender}, \citenamefont {Gulley}, \citenamefont
  {Jackman}, \citenamefont {Ullmann}, \citenamefont {Wolfsberg},\ and\
  \citenamefont {Nortier}}]{Mashnik2602}%
  \BibitemOpen
  \bibfield  {author} {\bibinfo {author} {\bibfnamefont {J.~W.}\ \bibnamefont
  {Weidner}}, \bibinfo {author} {\bibfnamefont {S.~G.}\ \bibnamefont
  {Mashnik}}, \bibinfo {author} {\bibfnamefont {K.~D.}\ \bibnamefont {John}},
  \bibinfo {author} {\bibfnamefont {F.}~\bibnamefont {Hemez}}, \bibinfo
  {author} {\bibfnamefont {B.}~\bibnamefont {Ballard}}, \bibinfo {author}
  {\bibfnamefont {H.}~\bibnamefont {Bach}}, \bibinfo {author} {\bibfnamefont
  {E.~R.}\ \bibnamefont {Birnbaum}}, \bibinfo {author} {\bibfnamefont {L.~J.}\
  \bibnamefont {Bitteker}}, \bibinfo {author} {\bibfnamefont {A.}~\bibnamefont
  {Couture}}, \bibinfo {author} {\bibfnamefont {D.}~\bibnamefont {Dry}},
  \bibinfo {author} {\bibfnamefont {M.~E.}\ \bibnamefont {Fassbender}},
  \bibinfo {author} {\bibfnamefont {M.~S.}\ \bibnamefont {Gulley}}, \bibinfo
  {author} {\bibfnamefont {K.~R.}\ \bibnamefont {Jackman}}, \bibinfo {author}
  {\bibfnamefont {J.~L.}\ \bibnamefont {Ullmann}}, \bibinfo {author}
  {\bibfnamefont {L.~E.}\ \bibnamefont {Wolfsberg}},\ and\ \bibinfo {author}
  {\bibfnamefont {F.~M.}\ \bibnamefont {Nortier}},\ }\href@noop {} {\bibfield
  {journal} {\bibinfo  {journal} {Applied Radiation and Isotopes}\ }\textbf
  {\bibinfo {volume} {70}},\ \bibinfo {pages} {2602} (\bibinfo {year}
  {2012})}\BibitemShut {NoStop}%
\bibitem [{\citenamefont {Griswold}\ \emph {et~al.}(2016)\citenamefont
  {Griswold}, \citenamefont {Medvedev}, \citenamefont {Engle}, \citenamefont
  {Copping}, \citenamefont {Fitzsimmons}, \citenamefont {Radchenko},
  \citenamefont {Cooley}, \citenamefont {Fassbender}, \citenamefont {Denton},
  \citenamefont {Murphy}, \citenamefont {Owens}, \citenamefont {Birnbaum},
  \citenamefont {John}, \citenamefont {Nortier}, \citenamefont {Stracener},
  \citenamefont {Heilbronn}, \citenamefont {Mausner},\ and\ \citenamefont
  {Mirzadeh}}]{Griswold16}%
  \BibitemOpen
  \bibfield  {author} {\bibinfo {author} {\bibfnamefont {J.~R.}\ \bibnamefont
  {Griswold}}, \bibinfo {author} {\bibfnamefont {D.~G.}\ \bibnamefont
  {Medvedev}}, \bibinfo {author} {\bibfnamefont {J.~W.}\ \bibnamefont {Engle}},
  \bibinfo {author} {\bibfnamefont {R.}~\bibnamefont {Copping}}, \bibinfo
  {author} {\bibfnamefont {J.~M.}\ \bibnamefont {Fitzsimmons}}, \bibinfo
  {author} {\bibfnamefont {V.}~\bibnamefont {Radchenko}}, \bibinfo {author}
  {\bibfnamefont {J.~C.}\ \bibnamefont {Cooley}}, \bibinfo {author}
  {\bibfnamefont {M.~E.}\ \bibnamefont {Fassbender}}, \bibinfo {author}
  {\bibfnamefont {D.~L.}\ \bibnamefont {Denton}}, \bibinfo {author}
  {\bibfnamefont {K.~E.}\ \bibnamefont {Murphy}}, \bibinfo {author}
  {\bibfnamefont {A.~C.}\ \bibnamefont {Owens}}, \bibinfo {author}
  {\bibfnamefont {E.~R.}\ \bibnamefont {Birnbaum}}, \bibinfo {author}
  {\bibfnamefont {K.~D.}\ \bibnamefont {John}}, \bibinfo {author}
  {\bibfnamefont {F.~M.}\ \bibnamefont {Nortier}}, \bibinfo {author}
  {\bibfnamefont {D.~W.}\ \bibnamefont {Stracener}}, \bibinfo {author}
  {\bibfnamefont {L.~H.}\ \bibnamefont {Heilbronn}}, \bibinfo {author}
  {\bibfnamefont {L.~F.}\ \bibnamefont {Mausner}},\ and\ \bibinfo {author}
  {\bibfnamefont {S.}~\bibnamefont {Mirzadeh}},\ }\href@noop {} {\bibfield
  {journal} {\bibinfo  {journal} {Applied Radiation and Isotopes}\ }\textbf
  {\bibinfo {volume} {118}},\ \bibinfo {pages} {366} (\bibinfo {year}
  {2016})}\BibitemShut {NoStop}%
\bibitem [{\citenamefont {Gauvin}(1963)}]{Gauvin}%
  \BibitemOpen
  \bibfield  {author} {\bibinfo {author} {\bibfnamefont {H.}~\bibnamefont
  {Gauvin}},\ }\href@noop {} {\bibfield  {journal} {\bibinfo  {journal}
  {Journal de Physique}\ }\textbf {\bibinfo {volume} {24}},\ \bibinfo {pages}
  {836} (\bibinfo {year} {1963})}\BibitemShut {NoStop}%
\end{thebibliography}
\end{document}